\documentclass[aps,preprint]{revtex4}
\usepackage{epsfig}
\begin{document}
\newcommand{\be}{\begin{eqnarray}}
\newcommand{\ee}{\end{eqnarray}}

\tighten
\preprint{DO-TH 03/19}
\vspace*{1cm}
\title
{Suppressing the Background Process to QED Compton Scattering for Delineating 
the Photon Content of the Proton}
\author{\bf A. Mukherjee}
\email{asmita@physik.uni-dortmund.de}
\author{C. Pisano}\email{pisano@harpo.physik.uni-dortmund.de}
\affiliation{ Institut f\"ur Physik, Universit\"at Dortmund, D 44221
Dortmund, Germany}
\date{\today\\[2cm]}

\begin{abstract}
We investigate the QED Compton process (QEDCS) in $e p \rightarrow e 
\gamma p$ and $e p \rightarrow e \gamma X$, together with the major 
background coming from the virtual Compton scattering (VCS), where 
the photon is emitted from the hadronic vertex. We suggest  new kinematical 
constraints which suppress the VCS  background and are furthermore suitable for 
the extraction of the equivalent photon content of the proton at the 
HERA collider. We show that the cross section, commonly expressed in terms 
of the proton structure functions, is reasonably well described by
the equivalent photon approximation of the proton, also in the  
inelastic channel in the proposed kinematical region. 
\end{abstract}
\maketitle

\section{Introduction}
QED Compton scattering (QEDCS) in the process $e p \rightarrow e \gamma X$, 
where $X$ is a general hadronic system, has long been suggested 
\cite{kessler,blu,rujula} as a unique possibility to determine the  photon 
content of the proton $\gamma(x,Q^2)$, evaluated in 
the equivalent photon approximation (EPA), which
is a very convenient and efficient tool to calculate cross sections having 
photon induced subprocesses \cite{kniehl,drees,ohn,gsv,gpr1,gpr2}.
Recently, the QED Compton process, depicted in Fig. 1, has been analyzed 
in \cite{lend} appropriate for measurements at HERA.  The corresponding events have a 
distinctive experimental signature: they are characterized by an electron 
and a photon in the final state, with their transverse momenta almost 
balancing each other and with little or no hadronic activity at the detector. 
In order to extract $\gamma(x,Q^2)$, several kinematical constraints have been
imposed which suppressed the major background contribution coming from the
virtual Compton scattering (VCS), depicted in Fig. 2, and also reduced the
contributions from initial and final state radiation effects \cite{thesis}
unrelated to QED Compton scattering. Although the cross section in
the elastic channel, $e p \rightarrow e \gamma p$, is very accurately
described by the equivalent photon approximation (EPA), substantial
discrepancy was observed in the inelastic channel and it was concluded that
the EPA does not give an accurate description of the process in this channel
\cite{lend}. In our previous paper \cite{pap1}, we did an independent study 
of the QED
Compton process, subject to the kinematical cuts of the HERA experiment, and
confirmed this result. We also showed that a measurement in bins of the
variable $x_\gamma$ shows better agreement with the EPA than the
corresponding measurements in bins of the leptonic variable $x_l$ (for
definitions, see section III). 

The purpose of this paper is to calculate the VCS background process, to 
study its relevance and to suggest new  cuts to be imposed on the
cross section for a more accurate extraction of $\gamma(x,Q^2)$. We
perform a detailed study of the full process $ep \rightarrow e \gamma
X$, both in the elastic (when $X$ is a proton) and inelastic channel, taking
into account the VCS, whose cross section in the inelastic channel is estimated utilizing an 
effective parton distribution of the
proton. In the elastic channel, to make a relative
estimate of the VCS, we take the proton to be pointlike and replace
the vertex by an effective vertex. We suggest new kinematical constraints to
suppress the inelastic VCS background, which turns out to be important in the
phase space domain of the HERA experiment. We also investigate the impact 
of these constraints  on the QEDCS cross section. We show that in the 
phase space region suggested by us and accessible at HERA, 
the EPA provides a reasonably good description of the QEDCS cross section, also in the inelastic channel.

The plan of the paper is as follows. In sections II and III, we present 
the cross
section in the elastic and inelastic channel respectively. Numerical results 
are 
given in section IV.
Summary and conclusions are presented in section V. The explicit form of the
matrix elements are shown in Appendices A and B.
  
\section{Elastic channel}
The cross section for the elastic process
\be
e(l)+p(P) \rightarrow e(l')+p(P')+\gamma(k')
\label{eq1}
\ee
can be written as \cite{pap1} 
\be
\sigma_{\mathrm{el}}(S) &= &{\alpha^3\over {2 (S-m^2)}} \int {d\hat s\over 2 \pi} \,dPS_2
(l+P;l'+k',P')\, dPS_2(l+k;l',k') \,\overline {{\mid {M_{\mathrm{el}}}\mid }^2},  
\label{elcross}
\ee
where we have defined the invariants
\be
S=(P+l)^2, ~~~~~~~~~\hat s=(l+k)^2, ~~~~~~~~t=k^2.
\label{invar}
\ee
$k=P-P'$ is the momentum transfer between the initial and the final proton 
and  $k'$ is the momentum of the final state observed (real) photon. 
As in \cite{pap1}, we neglect the
electron mass $m_e$ everywhere except when it is necessary to avoid divergences in
the formulae and take the proton to be massive, $ P^2=P'^2=m^2 $. The   
relevant Feynman diagrams for this process are shown in Fig. 1, with $X$
being a proton and $P_X \equiv P'$.

The Lorentz invariant $N$-particle phase-space element is written as
\be
dPS_N(P;P_1,...,P_N)=(2 \pi)^4 \delta \bigg (P-\sum_{i=1}^N P_i \bigg ) \prod_{i=1}^N
{d^3P_i\over { (2 \pi)^3 2 P_i^0}}.
\ee
We also write
\be
dPS_2(l+P;l'+k',P')={dt\over {8 \pi (S-m^2)}}
\label{lone}
\ee
and
\be
dPS_2(l+k;l',k')={d \hat t d\varphi^*\over {16 \pi^2 (\hat s-t)}}.
\label{ltwo}
\ee
Here $\varphi^*$ is the azimuthal angle of the outgoing $e-\gamma$ system in 
the $e-\gamma$ center-of-mass frame and $\hat t=(l-l')^2$.

Substituting Eqs. (\ref{lone}) and (\ref{ltwo}) in Eq. (\ref{elcross}) we get
\be
\sigma_{\mathrm{el}}(S)={\alpha^3\over {8 \pi (S-m^2)^2}} \int_{m_e^2}^{(\sqrt{S}-m)^2} d\hat s \int_{{t}_{\mathrm{min}}}^{t_{\mathrm{max}}}
dt \int_{\hat t_{\mathrm{min}}}^{\hat t_{\mathrm{max}}} d\hat t \int_0^{2\pi} d\varphi^* {1\over {(\hat s-t)}}
\overline {{\mid {M_{\mathrm{el}}}\mid }^2}.
\label{elsig}
\ee
The limits of integrations in Eq. (\ref{elsig}) follow from kinematics and
are given explicitly by Eqs. (2.18) and (2.24) in  \cite{pap1}. However, as
it will be discussed in section IV, we will impose additional  kinematical cuts
relevant to the experiment at HERA. \\ 
As already shown in \cite{pap1}
\be
\overline {{\mid {M_{\mathrm{el}}^{QEDCS}}\mid }^2} = \frac{1}{t^2}\, T_{\mu \nu}(l,k;l',k')\, H^{\mu \nu}_{\mathrm{el}} (P,P') ~,
\ee
where $T_{\mu \nu}$ is the leptonic tensor given by Eq. (2.9) of
\cite{pap1} divided by $e^4$, with $e$ denoting the electron charge. $H_{\mu \nu}$ is the hadronic tensor:
\be
H^{\mu \nu}_{\mathrm{el}}(P,P')=  [ H_1(t) (2 P-k)^\mu (2 P-k)^\nu +
H_2(t) (t g^{\mu
\nu}-k^\mu k^\nu)],
\ee
with
\be 
H_1(t)=\frac{G_E^2(t)- (t/4 \,m^2)\, G_M^2(t)}{1-{t/4\, m^2}},~~~~~~~~
H_2(t)=G_M^2(t).
\ee
The electric and magnetic form factors can be expressed as a combination 
of the real form factors $F_1(t), \, F_2(t)$:
\be
G_E(t)=F_1(t)- \tau F_2(t); ~~~~~~~~~~~G_M(t)=F_1(t)+F_2(t),
~~~~~~\tau={-t/4 m^2},
\ee
and they  are empirically parametrized
as dipoles: 
\be
G_E(t)= {1\over [1 - t/(0.71\,\mathrm{GeV}^2)]^{2}},~~~~G_M(t)=2.79~G_E(t).
\label{dipole}
\ee 
The full cross section for the process given by Eq. (\ref{elsig}) also
receives a contribution from the VCS in Fig. 2. 
The cross section for this process can be expressed in terms of off-forward or
generalized parton distributions \cite{ji}. In addition, there are
contributions due to the interference between the QEDCS and VCS.  In order to 
make a numerical estimate of these effects, one needs some 
realistic parametrization of the off-forward  distributions.
Our aim is to  estimate the VCS background so as to find the kinematical cuts
necessary to suppress it. We make a simplified
approximation to calculate the VCS cross section. We take the proton to be a
massive pointlike fermion, with the  equivalent $\gamma^* p$ vertex described
by a factor $-i\gamma^\mu F_1(t)$. 
Incorporating the background effects, the cross section of the process in Eq.
(\ref{eq1}) is given by Eq.(\ref{elsig}), where $\overline {{\mid {M_{\mathrm
{el}}}\mid }^2}$ now becomes
\be
\overline {{\mid {M_{\mathrm{el}}}\mid }^2}= \overline {{\mid {M^{QEDCS}_{\mathrm{el}}}\mid
}^2}
+\overline {{\mid {M^{VCS}_{\mathrm{el}}}\mid }^2}- 2\, {\Re{\bf{\it e}}}\,\overline {M^{QEDCS}_{\mathrm{el}}
M^{VCS *}_{\mathrm{el}}}.
\ee
The interference term will have opposite sign if we consider a positron
instead of an electron. 
The explicit expressions of $\overline {{\mid {M^{QEDCS}_{\mathrm{el}}}\mid }^2}$, 
$\overline
{{\mid {M^{VCS}_{\mathrm{el}}}\mid }^2}$ and 
$2\, \Re {\bf\it{e}} \,\overline {M^{QEDCS}_{\mathrm{el}}M^{VCS *}_{\mathrm{el}}}$ are given 
in appendix A.
The effect of proton mass is small in the kinematical range of HERA.

\section{Inelastic Channel}

We next consider the corresponding inelastic process, where an electron and 
a photon are produced in the final state  together with
a general hadronic system $X$:
\be
e(l)+p(P) \rightarrow e(l')+\gamma(k')+X(P_X),
\label{eqin}
\ee
with $P_X=\sum_{X_i} P_{X_i}$ being the sum over all momenta of the hadronic 
system $X$. The exact calculation of the
QEDCS rates follows our treatment in \cite{pap1} based on the ALLM97
parametrization \cite{allm} of the proton structure function $F_2(x_B, Q^2)$.

For the purpose of evaluating the relative importance of the VCS background 
we resorted to a unified parton model estimate of the VCS and QEDCS rates. 
The cross section within the parton model is given by
\be
\frac{d \sigma_{\mathrm{inel}}}{d x_B \,d Q^2 \,d \hat s\, d \hat t \,d \varphi^*}=\sum_{q} \, q(x_B,Q^2)\, \frac{d \hat
\sigma^q}{d \hat s\, d Q^2 \,d \hat t\, d \varphi^*},
\ee
where $q(x_B,Q^2)$  are the  quark and antiquark distributions of the initial 
proton, $q = u,\, d, \,s, \,\bar{u}, \,\bar{d}, \,\bar{s}$. Furthermore, $Q^2 
=- k^2 = -(l' + k' -l)^2$, $x_B= \frac{Q^2} {2 P \cdot (-k)}$  and  $d \hat \sigma^q$ is the 
differential cross section of the subprocess
\be
e(l)+q(p) \rightarrow e(l')+\gamma(k')+q(p').
\label{eqsub}
\ee
The relevant integrated cross section is given by 
\be 
\sigma_{\mathrm{inel}}(S)&=& 
{\alpha^3\over {8 \pi (S-m^2)^2}} \sum_q \int_{W^2_{\mathrm{min}}}^{W^2_{\mathrm{max}}}
dW^2 \int_{m_e^2}^{(\sqrt{S}-W)^2} d\hat s \int_{Q^2_{\mathrm{min}}}^{Q^2_{\mathrm{max}}}
 {dQ^2 \over Q^2}
\int_{{\hat t}_{\mathrm{min}}}^ {{\hat t}_{\mathrm{max}}}
d\hat t \int_0^{2\pi} d\varphi^* {1\over {(\hat s+Q^2)}}
\nonumber\\&&~~~~~~~~~~~~\times ~\overline {{\mid {M_{\mathrm{inel}}}
\mid }^2} ~~q(x_B,Q^2),
\label{insig}
\ee
with $W^2 = (p-k)^2 = m^2 + Q^2 \,(1-x_B)/x_B$. The limits of integration
are given explicitly by Eqs. (2.18), (3.11) and (3.12) of \cite{pap1} with
$\hat s_{min}=m_e^2$. 
Further constraints, related to the HERA kinematics, will be discussed in
the numerical section.   
Similar to the elastic channel, we have
\be 
\overline {{\mid {M_{\mathrm{inel}}}\mid }^2}= \overline {{\mid {M^{QEDCS}_{\mathrm{inel}}}\mid
}^2}
+\overline {{\mid {M^{VCS}_{\mathrm{inel}}}\mid }^2} -2\, {\Re{\it e}} \overline {M^{QEDCS}_{\mathrm{inel}}
M^{VCS *}_{\mathrm{inel}}}.
\ee 
Again, the interference term will have opposite sign for a positron.  
The explicit expressions are given in Appendix B. They are also given in
\cite{brodsky,metz} for a massless proton. We point out that the 
analytic expression of the QED Compton scattering
cross section in the inelastic channel was already given in \cite{pap1} in
terms of the proton structure functions $F_2(x_B,Q^2)$ and $F_1(x_B,Q^2)$. 

Furthermore, we introduce the auxiliary invariants $\hat S=(p'+k')^2$ 
and $\hat U= (p'-k)^2$, which can be written in terms of measurable 
quantities, 
\be
\hat S={\hat t (x_l-x_B)\over x_l}, ~~~~~~\hat U=\hat t- \hat S+Q^2,
\ee
with $x_l = {-\hat t\over 2 P \cdot (l-l')}$. In addition to the leptonic
variable $x_l$ we
define $x_\gamma={l\cdot k\over P\cdot l}$, which represents the fraction
of the longitudinal momentum of the proton carried by the virtual photon
\cite{pap1}. In the limit of the EPA, both $x_l$ and $x_\gamma$  are the same 
and become
equal to $x={\hat s\over S}$.
\section{Numerical Results}

In this section we present our numerical results.  In
order to select the QEDCS events, certain kinematical constraints are imposed 
in the Monte Carlo studies in \cite{thesis,
lend}. We introduce the following lab 
frame variables: energy of the final electron $E_e'$, energy of the final  
photon $E_{\gamma}'$,    
polar angles  of the outgoing electron and photon, $ \theta_e$ and   
$\theta_{\gamma}$ respectively,  and  acoplanarity angle $\phi$,     
which is defined as $\phi=|\,\pi-|\phi_{\gamma} -\phi_{e}|\,|$,     
where $\phi_{\gamma}$ and $\phi_{e}$ are the azimuthal angles of  
the outgoing photon and electron respectively ($0 \le \phi_{\gamma},\,
\phi_e \le 2 \,\pi$). The cuts are given in column A of Table I (from
hereafter, they will be referred to as the set A). The energies of the 
incoming particles are: $E_e = 27.5\,\,
\mathrm{GeV}$ (electron) and $E_p = 820 \,\,\mathrm{GeV}$ (proton).
So far the photon and  the electron in the final state have been identified 
only in the backward part of the H1 detector at HERA.  
To select signals where there 
are no hadronic activities near the two electromagnetic clusters,  the final 
hadronic state must not be found above the polar angle $\theta_h^{\mathrm{max}}
= \pi/2$ \cite{lend}.  
Motivated by this experimental arrangement, we have identified $\theta_h$ 
with the polar angle of the final quark 
$q'$ in the subprocess $e q \rightarrow e \gamma q'$.  It can be shown that
$\theta_h$ is  given by
\be
\cos \theta_h \equiv \cos\theta_{q'} = {1\over E_{q'}} (x_B E_p-E_e-E_e' 
\cos \theta_e-E_\gamma'
\cos\theta_\gamma)
\ee
and  $E_{q'}  = x_B E_p+E_e-E_e'-E_\gamma'$  being the 
energy of the 
final parton. Here we have assumed that the final hadrons are emitted 
collinearly with the struck quark $q'$. For the elastic process 
$\theta_h \equiv \theta_{p'}$, the polar angle of the scattered proton, 
can be obtained
by substituting $x_B=1$ in the above expression. 
Thus we impose the additional condition \cite{lend}
\be
\theta_h < \pi /2
\label{cut4}
\ee
on the cross section. However, no constraint on the hadronic final state was
used in the cross section calculation presented in \cite{lend}. 
Inclusion of Eq. (\ref{cut4}) reduces the QEDCS cross section by about $10\%$.


 In the
kinematical region defined by the constraints mentioned above, the contributions from the 
initial and final state 
radiation, unrelated to QED Compton scattering, are suppressed 
\cite{kessler,blu,rujula,thesis}. 
Furthermore, we checked that the event rates related to the elastic VCS 
process and  its interference with elastic QEDCS are negligible compared 
to the ones
corresponding to pure elastic QEDCS. This is expected because the elastic QEDCS cross 
section is very much dominated by the small values of the variable $-t$, 
compared to $-\hat t$, see Eqs. (\ref{qel}) and (\ref{vel}).  
Such an observation is similar to that of
\cite{thesis}, where the elastic DVCS background was calculated using a Regge
model in different kinematical bins.    
Our estimate was done  taking the 
proton to be pointlike with an effective vertex, as discussed in section II. 
We find that, in this approximation, the elastic QEDCS cross section differs 
from the actual one in \cite{pap1} by about $3 \% $ within the range defined by
the kinematical constraints.  

Fig. 3 shows the total (elastic + inelastic) QEDCS cross section in 
$x_l-Q_l^2$ bins with $Q_l^2=-\hat t$, subject to the cuts of set A. 
For comparison we have also plotted the cross section without the cut 
on $\theta_h$, similar to our analysis in \cite{pap1}. This additional 
constraint affects the result only in the inelastic channel.

We  checked that the upper limit in Eq. (\ref{cut4})  reduces the contribution 
from the inelastic VCS reaction. In order to calculate it, one needs a model 
for the parton distributions $q(x_B,Q^2)$.
However, in the relevant kinematical region, $Q^2$ can be very small and
may become close to zero, where the parton picture is not applicable.
Therefore, in our estimate, we replace the parton distribution
$q(x_B,Q^2)$ by an effective parton distribution 
\be
\tilde{q}(x_B,Q^2)={Q^2\over Q^2+a\, Q_0^2}\, q(x_B,Q^2+Q_0^2),
\label{fq}
\ee
where $a=1/4$ and $Q_0^2=0.4 \, \mathrm{GeV}^2$  are two parameters and 
$q(x_B,Q^2)$ is the NLO GRV98 \cite{grv} parton distribution. 
$Q_0^2$ prevents the scale in the distribution to become too 
low. Eq. (\ref{fq}) is  motivated by a similar form used in \cite{bad,allm} 
for the parametrization of the structure function $F_2(x_B,Q^2)$ in the low 
$Q^2$ region. It is
clear that at high $Q^2$, $\tilde{q}(x_B,Q^2) \rightarrow q(x_B,Q^2)$.

In this paper, we introduce a new set of cuts, which are given in the
column B of Table \ref{tableone} (and will be referred to as the set B) for a 
better extraction of the equivalent photon distribution of the proton as 
well as to suppress the VCS background. These cuts
will be compared to the set A in the following. Instead of the constraint 
on the acoplanarity, namely $\phi < \pi/4$,
where the upper limit is actually ambiguous, we impose $\hat s > Q^2$. The relevance of the cut $\hat S \gtrsim \hat s $  can be seen from Fig.
4. This shows the cross sections of  the QEDCS and VCS processes 
in the inelastic channel, calculated using Eq. (\ref{insig}) and subject to the
kinematical limits of set B (except $ \hat S \gtrsim \hat s $), 
in bins of $\hat s-\hat S$.
Fig. 4 shows that the VCS cross section is higher than QEDCS
for bins with $\hat s \gtrsim \hat S$  but falls sharply in bins
for which $\hat s $ is close to $\hat S$ and becomes much suppressed for
$\hat S \gtrsim \hat s$ . This is expected because $\hat S$ corresponds to the
quark propagator in the VCS cross section, see Eq. (\ref{vin}), and a lower 
value enhances this
contribution. In fact the sharp drop of the VCS cross section in bins where 
$\hat S \gtrsim \hat s$ is due to the fact that both the propagators 
$\hat s, \,\hat u$ in the QEDCS cross section are constrained to be 
smaller than 
$\hat S, \,\hat U$ for  VCS in these bins, see Eqs. (\ref{qin}), (\ref{vin}). 
The QEDCS cross section is always
enhanced by the factor  $Q^2$ in the denominator of Eq. (\ref{qin}) coming 
from the virtual photon,  which can be very small in the
kinematical region of interest here. This plot shows that \underline{imposing} 
\underline{a cut on $\hat S$ can be very effective in reducing the background 
contribution from VCS}. The interference between inelastic QEDCS
and VCS gives negligible contribution. We have also shown the QEDCS cross
section using the ALLM97 parametrization of $F_2(x_B,Q^2)$ \cite{pap1}. 
The discrepancy between this and the one calculated using the
parametrization in Eq. (\ref{fq}) is less than $5 \%$ in almost all the bins, and 
maximally $7\%$ in two bins.  

In Fig. 5(a), we have shown the inelastic QEDCS and VCS cross 
sections in bins
of $x_\gamma$, subject to the cuts of set A. The VCS
cross section is much suppressed in the smaller $x_\gamma$ bins but becomes
enhanced as $x_\gamma$ increases, which indicates that such a set of cuts is 
\underline{not}
suitable  to remove the background at higher $x_\gamma$. The situation will
be the same in $x_l$ bins. Fig. 5(b) shows the
cross sections but with the set B. The background in this case is suppressed 
for all $x_\gamma$ bins,  which means that such a cut is more effective in 
extracting QEDCS
events also for higher $x_\gamma$. In addition, we have plotted the QEDCS
cross section in terms of the structure function $F_2(x_B, Q^2)$, using 
the ALLM97 parametrization. Fig. 5 shows that our parametrizaton gives 
a reasonably good description of the proton, at least for the QEDCS process, 
in most of the bins except those with high $x_\gamma$. However, 
this parametrization has been used only to make a relative estimate of the 
 background events. In fact, a quantitative estimate of the inelastic VCS
events has not been presented in \cite{thesis,lend}. 

Figs. 6(a) and 6(b) show the  QEDCS cross section in bins of $x_l$ and 
$x_\gamma$, respectively, subject to 
the constraints of set B. The elastic cross section has been calculated using 
Eqs. (\ref{elsig})-(\ref{dipole}), as in
\cite{pap1}. The inelastic cross section is given by Eq.
(3.10) of \cite{pap1} in terms of the structure functions $F_1(x_B, Q^2)$ 
and $F_2(x_B, Q^2)$.
We have assumed the Callan-Gross relation and used the ALLM97 parametrization
\cite{allm} for $F_2(x_B,Q^2)$. In this way the results presented in Fig. 
6, labelled as 'exact', are free from the parton model approximations 
in Figs. 4 and 5. 
In the same plot, we have also 
shown the total cross section calculated in terms of the EPA,
according to Eqs. (2.28) and (3.13) in \cite{pap1}. 
Fig. 6(b)  shows much better agreement between the approximate 
cross section based on the EPA and the 'exact' one.
For Fig. 6(a), the discrepancy 
is about $3-7 \%$ in
the first three bins, between $20-30 \%$ in three other bins and higher 
in the last bin. 
In Fig. 6(b) it is
$1-6 \%$ in five  bins, $13-15 \%$ in two bins and about $25 \%$ in the 
last bin. 
The discrepancy of the 'exact' cross section, integrated over $x_{\gamma}$, 
with the approximate one, when subject to the 
constraints  of set B is $0.38\%$ in the elastic channel and $4.5\%$ 
in the inelastic one. The total (elastic + inelastic) discrepancy turns out to
be $2.26\%$, which should be compared to the  values $14 \%$,  already
observed in \cite{pap1} when subject to the set A, except the one on
$\theta_h$, and  $24 \%$ when this one is imposed too. 

As we know, the elastic QEDCS cross section is described very accurately 
by the EPA \cite{thesis,pap1}. It is thus more interesting to investigate
the inelastic channel in this context.  
The elastic QEDCS events can be separated from the inelastic ones by
applying a cut on $\theta_h$. We have found that, with the restriction
$\theta_h \ge 0.1^\circ$, the elastic events are rejected and 
all the inelastic events are retained in the cross section. A lower limit on
$\theta_h $ higher than $1^\circ$ removes a substantial part (more than 
$30\%$) of the inelastic events.   

Table \ref{tabletwo} shows the 'exact' inelastic QEDCS cross section  
in $x_l$ and $Q_l^2$ bins, subject to the cuts A. 
We have also shown the cross section in
the EPA with the same constraints (the last two cuts of set A are not relevant in this case). The discrepancy with the EPA is quite substantial. We have also
shown the results with the cuts B, both the 'exact' and the one in terms of 
the EPA, in the
same table (the constraint $\hat s > Q^2$ is not relevant for the EPA). The
discrepancy between the 'exact' and the EPA here is much less and on the
average it is $20 \%$. Table \ref{tablethree} is almost similar, the only 
difference is that
the bins are now in $x_\gamma$. With the cuts of set A, the discrepancy now
is on the average $50 \% $, whereas, with the cuts B, the average
discrepancy is $17 \%$.    
   
Our results  show that the extraction of the equivalent photon
distribution $\gamma(x,Q^2)$ is very much dependent on the kinematical 
constraints utilized to single out QEDCS events, in particular on the 
one on acoplanarity. The kinematical limits  presented here are much more appropriate than
those suggested in \cite{thesis} 
for a reliable extraction of $\gamma(x,Q^2)$.  It is also clear that 
this  discrepancy is  entirely due to  the inelastic channel, which was also
observed in  \cite{thesis,pap1}.

\section{summary and conclusions}
To summarize, in this paper we have analyzed the QED Compton process,
relevant for the experimental determination of the equivalent photon
distribution of the proton $\gamma(x,Q^2)$. We have also calculated the major
background process, namely virtual Compton scattering, assuming an effective
parametrization of the parton distributions of the proton, both in 
the elastic and inelastic channels.
The elastic VCS is suppressed compared to the QEDCS, in the phase space 
region accessible at HERA. We have shown that a constraint on the
invariants $\hat S \gtrsim \hat s $ is very effective in removing the 
inelastic VCS background. Furthermore, the
selection of the QEDCS events in the process $e p \rightarrow e \gamma X$
is sensitive to the specific kinematical limits, in particular to the upper
limit  of the acoplanarity angle $\phi$, which was used in the recent 
analysis \cite{lend,thesis} of events as observed with the HERA-H1 detector. 
Instead of the acoplanarity, one can
also directly impose cuts on the invariants, like $\hat s > Q^2$ (both of
them are measurable quantities), which directly restricts one to the
range of validity of the EPA. With these constraints, the total (elastic +
inelastic) cross section agrees with the EPA within $3 \%$. Thus, we conclude 
that by choosing the kinematical domain relevant for this approximation 
carefully, it is possible
to have a more accurate extraction of $\gamma(x,Q^2)$. This will also give
the region of validity of the EPA, which is important to have a convenient 
and reliable estimate of the photon induced subprocesses
in $ep$ and $pp$ colliders.             

\section{Acknowledgements}

We warmly acknowledge E. Reya and M. Gl\"uck for initiating  this study, as
well as for  many helpful discussions and suggestions. We also thank W.
Vogelsang and V. Lendermann for
helpful discussions. This work has been supported in part by
the 'Bundesministerium f\"ur Bildung und Forschung', Berlin/Bonn. 
   
\appendix
\section{Matrix element for the elastic process}
In this Appendix, we give the expressions of $\overline {{\mid 
{M^{QEDCS}_{\mathrm{el}}}\mid }^2 }$, $\overline {{\mid 
{M^{VCS}_{\mathrm{el}}}\mid }^2} $ and $2\, {\Re\it{e}}\, \overline {M^{QEDCS}_{\mathrm{el}}
M^{VCS *}_{\mathrm{el}}}$ 
corresponding to Eq. (\ref{elsig}):

\be 
\overline {{\mid {M^{QEDCS}_{\mathrm{el}}}\mid }^2} = \frac {4}{t ~\hat s~\hat u} ~\bigg [{A +  \frac{2 m^2}{t} B}\bigg ]
~F_1^2(t) ,
\label{qel}
\ee
\be 
\overline {{\mid {M^{VCS}_{\mathrm{el}}}\mid }^2 } =  \frac{4}{\hat t\, U'\, \hat S'}
 \, \bigg [ A - \frac{2 m^2}{\hat t~ U'~ \hat S'}\, C \bigg ] ~ F_1^2(\hat t) ,
\label{vel}
\ee
with $ \hat S = -(\hat s + \hat u + U' - m^2) $, $\hat S' = \hat S-m^2$ and
\be
A&=& 2~t^2 -    2~t~(\hat s - 2~S' - U')  
+ \hat s^2 - 2~\hat s~S' \nonumber \\ && +\,\, 4~S'^2 + 2~S'~\hat u + \hat u^2
+ 4~S'~U' + 2~\hat u~U' + 2~U'^2,
\ee
\be
B & = & 2~ t^2 - 2 ~t ~(\hat s + \hat u) + \hat s^2 + \hat u^2,
\ee
\be
C&=& (\hat s + \hat u)^2~[t^2 + \hat s^2 - 2~t~(\hat s - S') - 2~\hat s~S'
\nonumber\\&&+ 
      2~S'^2 + 2~S'~\hat u + \hat u^2 - 2~m^2~( \hat s + \hat u -t)] 
   \nonumber\\&&+ 2~(\hat s + \hat u)~[t^2 -t~\hat s + \hat u~(-\hat s + 2~S' + \hat u)]~U' \nonumber\\&& + 2~[t^2 + \hat u^2 -t~(\hat s + \hat u)]~U'^2.
\ee
We have introduced the invariants $ U=(P-k')^2, ~\hat u=(l-k')^2 $
and used the notations $S'=S-m^2, ~U'=U-m^2$  for
compactness.

For the interference between QEDCS and VCS  we have
\be
2\, {\Re\it{e}} \,\overline {M^{QEDCS}_{\mathrm{el}}M^{VCS *}_{\mathrm{el}}}=4\,\frac{D +2m^2E}
{ t~\hat s~\hat u~\hat t~U'~\hat S'}~F_1(\hat t) F_1(t),
\ee
with

\be
D&=&\{ (\hat s + \hat u)~[t~\hat u + S'~(\hat s + \hat u)] + 
   [\hat s~(\hat s + \hat u) -t~(\hat s - \hat u)]~U'\}~ \nonumber \\&&~~~
[2~t^2 
+ \hat s^2 -2~\hat s~S' +  
   4~S'^2 + 2~S'~\hat u + \hat u^2 - 2~t~(\hat s - 2~S' - U')\nonumber\\ && 
~~~~~~~~~~~~+ 4~S'~U' + 2~\hat u~U' + 2~U'^2],
\ee
\be
E&=& -S'~\hat u^3 - \hat s^3~(S' - 2~\hat u + U') - 
   \hat s^2~\hat u~(7~S' + 2~U') -   \hat s~\hat u^2~(7~S' + 2~\hat u + 5~U')
  \nonumber \\ && ~~~+ 
   2~t^2~[\hat s~(\hat u - U') + \hat u~(\hat u + U')] - 
   t~ (\hat s + \hat u)~[ \hat s~(-2~S' + 3~\hat u - 3~U') \nonumber \\ 
&&~~~~~~~+ \hat u~(-2~S' + \hat u + U')].
\ee

\section{Matrix element for the inelastic process}
Here we give the expressions of $\overline {{\mid {M^{QEDCS}_{\mathrm{inel}}}
\mid}^2}$, $\overline {{\mid {M^{VCS}_{\mathrm{inel}}}\mid }^2}$
and $2 \,{\Re \it{e}}\,\overline {M^{QEDCS}_{\mathrm{inel}}M^{VCS *}_{\mathrm{inel}}}$ corresponding to Eq.
(\ref{insig}):
\be
\overline {{\mid {M^{QEDCS}_{\mathrm{inel}}}\mid }^2}&=& -4\, e_q^2 \,\frac{F}{ Q^2\, \hat s \,\hat
u}~ ,
\label{qin}
\ee
\be
\overline {{\mid {M^{VCS}_{\mathrm{inel}}}\mid }^2}&=&4 \,e_q^4\,\frac{F} {\hat t ~\hat U~\hat S}~ ,
\label{vin}
\ee 
with $\hat S = -(\hat s + \hat u + x_B\, U')$,\, $ \hat U = x_B\,U'$ and
\be
F &=&  \hat s^2 + \hat u^2 + 2 \,\{ Q^4+ Q^2~[\hat s - (2~S' +
U')~x_B] + x_B\, (S'~\hat u + \hat u~U'\nonumber\\&&- \hat s~S') + x_B^2\,(
   2~S'^2 + 2~S'~U' + U'^2)\}.
\ee
Here $e_q $ is the charge of the parton in units of the charge of the
proton. Also we have
\be
2 \,{\Re\it{e}}\,\overline {M^{QEDCS}_{\mathrm{inel}}M^{VCS *}_{\mathrm{inel}}} &=&- 4\,e_q^3\,\frac{G}{Q^2~\hat s~\hat u~\hat t~\hat U~\hat S}
\ee
with 
\be
G & =& \{-Q^2~\hat u~(\hat s + \hat u) 
+ Q^2~(\hat s - \hat u)~U'~x_B + 
   (\hat s + \hat u)~[S'~\hat u \nonumber\\&&+ \hat s~(S' + U')]~x_B\}~\{2~Q^4 + \hat s^2 + \hat u^2 - 
   2~\hat s~S'~x_B + 2~Q^2~[\hat s - (2~S' \nonumber\\&&+ U')~x_B]+ 
   2~x_B~[\hat u~(S' + U') + (2~S'^2 + 2~S'~U' + U'^2)~x_B]\}.
\ee
The analytic form of the interference term agrees with \cite{brodsky} 
but differs from \cite{metz}
in the massless case slightly, in particular in Eq. (15) of \cite{metz}, $8$
in the first line should be replaced by $4$ and $(-8)$ in the sixth line
should be replaced by $(-16)$. However we have checked that this does not 
affect our numerical results for HERA kinematics.  
 

\newpage
\begin{table}[c]
\begin{center}
\vspace*{8cm}
\hspace*{-0.5cm}
\begin{tabular}{|c|c|}
\hline 
$ A $ & $B$ \\
\hline\hline

$~~E_e', \, E_{\gamma}' > 4 \, \mathrm{GeV}~~ $ & $~~E_e', \, E_{\gamma}' > 4 \,\mathrm{GeV}~~ $ \\
$E_e' + E_{\gamma}' > 20 \, \mathrm{GeV}$ & $ E_e' + E_{\gamma}' > 20\, \mathrm{GeV}$ \\
$~~~ 0.06 < \theta_e, \, \theta_{\gamma} < \pi - 0.06~~~$ & 
$~~~ 0.06 < \theta_e, \, \theta_{\gamma} < \pi - 0.06~~~$ \\
$ \phi < {\pi}/ {4}   $  & $ \hat s > Q^2 $ \\
$ \theta_h < {\pi}/{2}  $  & $ \hat S > \hat s $ \\

\hline

\end{tabular}
\end{center}
\caption{A: cuts to simulate HERA-H1 detector. B: cuts introduced in this paper.}
\label{tableone}

\end{table}


\begin{table}
\begin{center}
\hspace*{-0.5cm}
\begin{tabular}{|c|c|c|c||c|c|c|}
\hline 
$x_l$ bin & $Q^2_l$ bin & $\sigma_{\mathrm{inel}}$  & 
$\sigma_{\mathrm{inel}}^{\mathrm{EPA}}$ &$\sigma_{\mathrm{inel}}^*$ &
$\sigma_{\mathrm{inel}}^{\mathrm{EPA *}} $    \\ 
\hline \hline
          &           &             &     &      &  \\
$~1.78\times 10^{-5}-5.62 \times 10^{-5}~$ & $1.5 -2.5$ &$~5.697\times 10^1~$  & $1.529 \times 10^2 $ & $~1.097\times 10^2~$ & $ ~1.344\times 10^2~$    \\ 
$1.78\times 10^{-5}-5.62 \times 10^{-5}$ & $2.5 - 3.5$&$2.074\times 10^1$ &  $~3.362\times 10^1~$ & $4.067\times 10^1$  & $~2.994\times 10^1~$   \\ \hline
$5.62\times 10^{-5}-1.78\times 10^{-4} $ & $1.5 - 5.0$&$1.781\times 10^2$  & $4.116\times 10^2$  & $3.050\times 10^2$   & $3.518\times 10^2$ \\
$5.62\times 10^{-5}-1.78 \times 10^{-4}$&$ 5.0-8.5$&$8.681\times 10^1 $ &  $2.098\times 10^2$ & $1.467\times 10^2 $ &$1.847\times 10^2$  \\ 
$5.62 \times 10^{-5}-1.78\times 10^{-4}$ & $8.5-12.0$&$2.713\times 10^1 $&$8.091\times 10^1 $ & $4.523\times 10^1$  & $7.223\times 10^1$ \\ \hline
$1.78\times 10^{-4}-5.62\times 10^{-4}$ & $3.0-14.67 $ & $1.701\times 10^2$  &  $ 2.210\times 10^2$ & $2.464 \times 10^2$  &  $1.826\times 10^2$\\ 
$1.78\times 10^{-4}-5.62\times 10^{-4}$ & $14.67-26.33 $ & $8.057\times 10^1$ &$ 1.557\times 10^2$ & $1.264\times 10^2$ &  $1.363\times 10^2$\\ 
$1.78\times 10^{-4}-5.62\times 10^{-4}$ & $26.33-38.0 $ & $2.396\times 10^1$   &$ 4.558\times 10^1$ & $3.778\times 10^1$ & $4.017\times 10^1$ \\ \hline
$5.62\times 10^{-4}-1.78\times 10^{-3}$ & $10.0-48.33 $ & $9.102\times 10^1$ & $ 8.092\times 10^1$ & $1.081\times 10^2$  & $6.516\times 10^1$ \\ 
$5.62\times 10^{-4}-1.78\times 10^{-3}$ & $48.33-86.67 $ & $4.036\times 10^1$  & $ 5.272\times 10^1$  & $6.137\times 10^1$ &  $4.541\times 10^1$\\ 
$5.62\times 10^{-4}-1.78\times 10^{-3}$ & $86.67-125.0 $ & $1.154\times 10^1 $& $1.587\times 10^1 $ & $1.803\times 10^1$  & $1.378\times 10^1$ \\ \hline
$1.78\times 10^{-3}-5.62\times 10^{-3}$ & $22-168 $ & $4.282\times 10^1$ & $ 3.080\times 10^1$ & $4.272\times 10^1$ & $2.390\times 10^1$ \\ 
$1.78\times 10^{-3}-5.62\times 10^{-3}$ & $168-314 $ & $1.800\times 10^1$ & $ 2.059\times 10^1$  &$2.599\times 10^1$ & $1.752\times 10^1$ \\ 
$1.78\times 10^{-3}-5.62\times 10^{-3}$ & $314-460 $ & $6.467$ & $ 1.021\times 10^1$ &$8.928$ & $8.804$ \\ \hline
$5.62\times 10^{-3}-1.78\times 10^{-2}$ & $0-500 $ & $1.406\times 10^1$ &  $8.823 $ & $1.133\times 10^1$  & $6.048$ \\ 
$5.62\times 10^{-3}-1.78\times 10^{-2}$ & $500-1000 $ & $1.151\times 10^1$ & $ 1.687\times 10^1$ & $1.484\times 10^1$ & $1.425 \times 10^1$  \\ 
$5.62\times 10^{-3}-1.78\times 10^{-2}$ & $1000-1500  $ & $2.985 $ & $ 4.885$ 
& $3.708$  &  $4.090$ \\ \hline
$1.78\times 10^{-2}-5.62\times 10^{-2}$ & $0-1500 $ & $3.506$ & $1.811$  & $2.200$ & $1.030$ \\ 
 $1.78\times 10^{-2}-5.62\times 10^{-2}$ & $1500-3000 $ & $3.621$ & $ 4.867$
 & $4.139$ &  $3.908$ \\ 
$1.78\times 10^{-2}-5.62\times 10^{-2}$ & $3000-4500  $ & $9.366\times 10^{-1}$ &  $1.341$ & $1.028$ & $1.044$ \\ \hline
$5.62\times 10^{-2}-1.78\times 10^{-1}$ & $10-6005 $ & $1.079$ & $ 7.147\times 10^{-1}$ & $6.723\times 10^{-1}$ & $3.990 \times 10^{-1}$  \\ 
$5.62\times 10^{-2}-1.78\times 10^{-1}$ & $6005-12000 $ & $5.382\times 10^{-1}$  & $5.922\times 10^{-1}$  & $~4.953\times 10^{-1}~$ &  $3.890\times 10^{-1}$ \\ 
 $~ 5.62\times 10^{-2}-1.78\times 10^{-1} ~$  & $~12000-17995~$ & $~6.035\times 10^{-2}~ $& $~6.789\times 10^{-2}~$ & $~4.613\times 10^{-2}~$& $~3.662\times 10^{-2}~$   \\ 
     &    &          &        &         &    \\

\hline

\end{tabular}
\end{center}
\caption{Double differential QED Compton scattering cross section
(inelastic) in $x_l$ and $Q_l^2$ bins. $\sigma_{\mathrm{inel}}$ and 
$\sigma_{\mathrm{inel}}^*$
correspond to the 'exact' (without the EPA)  cross section  subject to the 
cuts A and B of Table I
respectively. $\sigma_{\mathrm{inel}}^{\mathrm{EPA}}$ and  $\sigma_{\mathrm{inel}}^{\mathrm{EPA
*}}$  correspond to the one in the EPA and  subject to the cuts A and B
respectively. $Q^2_l$ is expressed in 
${\mathrm{GeV}^2}$ and the cross-sections are in pb.}
\label{tabletwo}
\end{table} 

\newpage
\begin{table}
\begin{center}
\hspace*{-0.5cm}
\begin{tabular}{|c|c|c|c||c|c|c|}
\hline 
$x_{\gamma}$ bin & $Q^2_l$ bin & $\sigma_{\mathrm{inel}}$  & 
$\sigma_{\mathrm{inel}}^{\mathrm{EPA}}$ &$\sigma_{\mathrm{inel}}^*$ &
$\sigma_{\mathrm{inel}}^{\mathrm{EPA *}} $    \\ 
\hline \hline
          &           &             &     &      &  \\
$~1.78\times 10^{-5}-5.62 \times 10^{-5}~$ & $1.5 -2.5$ &$~5.331\times 10^1~$  & $1.529 \times 10^2 $ & $~1.022\times 10^2~$ & $ ~1.344\times 10^2~$    \\ 
$1.78\times 10^{-5}-5.62 \times 10^{-5}$ & $2.5 - 3.5$&$2.957\times 10^1$ &  $~3.362\times 10^1~$ & $5.368\times 10^1$  & $~2.994\times 10^1~$   \\ \hline
$5.62\times 10^{-5}-1.78\times 10^{-4} $ & $1.5 - 5.0$&$1.825\times 10^2$  & $4.116\times 10^2$  & $3.111\times 10^2$   & $3.518\times 10^2$ \\
$5.62\times 10^{-5}-1.78 \times 10^{-4}$&$ 5.0-8.5$&$1.151\times 10^2 $ &  $2.098\times 10^2$ & $1.856\times 10^2 $ &$1.847\times 10^2$  \\ 
$5.62 \times 10^{-5}-1.78\times 10^{-4}$ & $8.5-12.0$&$4.809\times 10^1 $&$8.091\times 10^1 $ & $7.550\times 10^1$  & $7.223\times 10^1$ \\ \hline
$1.78\times 10^{-4}-5.62\times 10^{-4}$ & $3.0-14.67 $ & $1.056\times 10^2$  &  $ 2.210\times 10^2$ & $1.523 \times 10^2$  &  $1.826\times 10^2$\\ 
$1.78\times 10^{-4}-5.62\times 10^{-4}$ & $14.67-26.33 $ & $9.862\times 10^1$ &$ 1.557\times 10^2$ & $1.432\times 10^2$ &  $1.363\times 10^2$\\ 
$1.78\times 10^{-4}-5.62\times 10^{-4}$ & $26.33-38.0 $ & $3.819\times 10^1$   &$ 4.558\times 10^1$ & $5.539\times 10^1$ & $4.017\times 10^1$ \\ \hline
$5.62\times 10^{-4}-1.78\times 10^{-3}$ & $10.0-48.33 $ & $4.717\times 10^1$ & $ 8.092\times 10^1$ & $5.829\times 10^1$  & $6.516\times 10^1$ \\ 
$5.62\times 10^{-4}-1.78\times 10^{-3}$ & $48.33-86.67 $ & $4.865\times 10^1$  & $ 5.272\times 10^1$  & $6.648\times 10^1$ &  $4.541\times 10^1$\\ 
$5.62\times 10^{-4}-1.78\times 10^{-3}$ & $86.67-125.0 $ & $1.774\times 10^1 $& $1.587\times 10^1 $ & $2.463\times 10^1$  & $1.378\times 10^1$ \\ \hline
$1.78\times 10^{-3}-5.62\times 10^{-3}$ & $22-168 $ & $2.222\times 10^1$ & $ 3.080\times 10^1$ & $2.452\times 10^1$ & $2.390\times 10^1$ \\ 
$1.78\times 10^{-3}-5.62\times 10^{-3}$ & $168-314 $ & $2.128\times 10^1$ & $ 2.059\times 10^1$  &$2.761\times 10^1$ & $1.752\times 10^1$ \\ 
$1.78\times 10^{-3}-5.62\times 10^{-3}$ & $314-460 $ & $8.593$ & $ 1.021\times 10^1$ &$1.131\times 10^1$ & $8.804$ \\ \hline
$5.62\times 10^{-3}-1.78\times 10^{-2}$ & $0-500 $ & $6.944$ &  $8.823 $ & $6.344$  & $6.048$ \\ 
$5.62\times 10^{-3}-1.78\times 10^{-2}$ & $500-1000 $ & $1.243\times 10^1$ & $ 1.687\times 10^1$ & $1.514\times 10^1$ & $1.425\times 10^1$  \\ 
$5.62\times 10^{-3}-1.78\times 10^{-2}$ & $1000-1500  $ & $3.572 $ & $ 4.885$ 
& $4.311$  &  $4.090$ \\ \hline
$1.78\times 10^{-2}-5.62\times 10^{-2}$ & $0-1500 $ & $1.568$ & $1.811$  & $1.101$ & $1.030$ \\ 
 $1.78\times 10^{-2}-5.62\times 10^{-2}$ & $1500-3000 $ & $3.720$ & $ 4.867$
 & $4.052$ &  $3.908$ \\ 
$1.78\times 10^{-2}-5.62\times 10^{-2}$ & $3000-4500  $ & $1.057$ &  $1.341$ & $1.121$ & $1.044$ \\ \hline
$5.62\times 10^{-2}-1.78\times 10^{-1}$ & $10-6005 $ & $6.448\times 10^{-1}$ & $ 7.147\times 10^{-1}$ & $4.548\times 10^{-1}$ & $3.990\times 10^{-1}$  \\ 
$5.62\times 10^{-2}-1.78\times 10^{-1}$ & $6005-12000 $ & $5.671\times 10^{-1}$  & $5.922\times 10^{-1}$  & $5.003\times 10^{-1}$ &  $3.890\times 10^{-1}$ \\ 
 $ 5.62\times 10^{-2}-1.78\times 10^{-1} $  & $~12000-17995~$ & $~7.343\times 10^{-2}~ $& $~6.789\times 10^{-2}~$ & $~5.664\times 10^{-2}~$& $~3.662\times 10^{-2}~$   \\ 
     &    &          &        &         &    \\

\hline

\end{tabular}
\end{center}
\caption{As in table II but for $x_{\gamma}$ bins.}
\label{tablethree}
\end{table} 


\newpage
\vspace*{.2cm}
\begin{center}
\epsfig{figure= 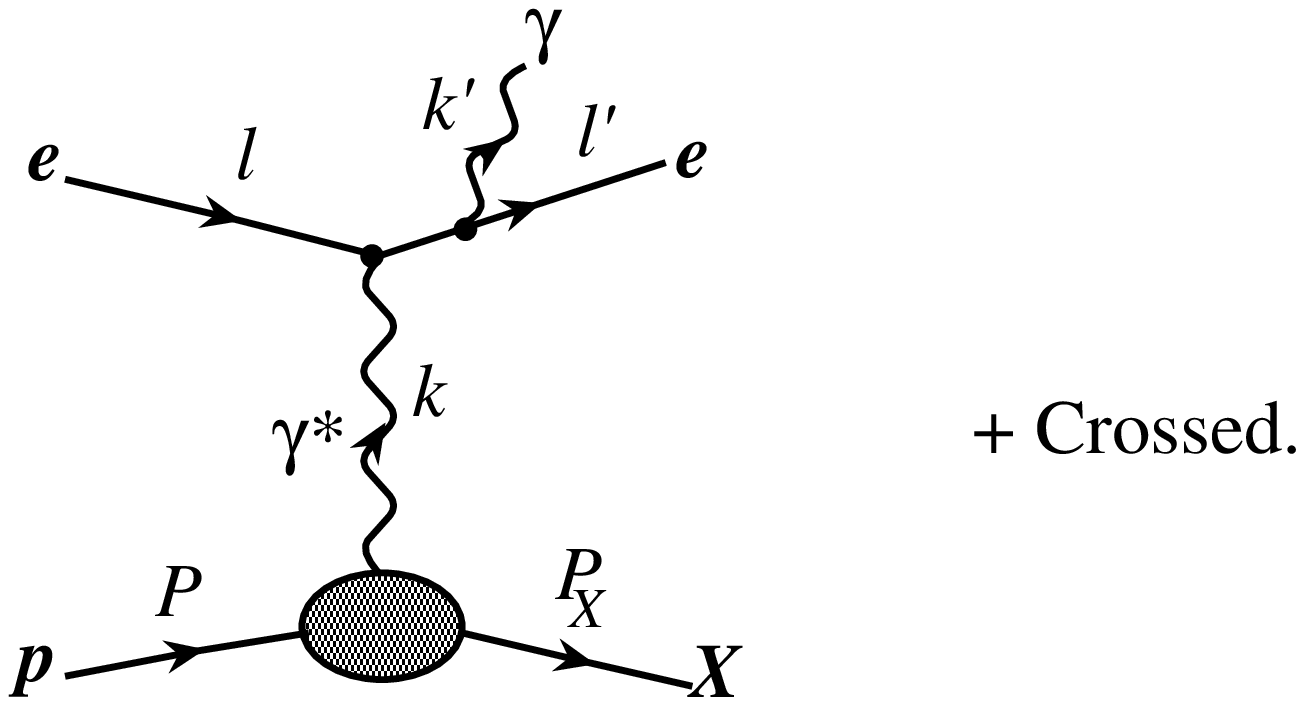, width=16cm, height= 6cm}\\
\end{center}
\begin{center}
\parbox{10.0cm}
{{\footnotesize
 Fig. 1:  Feynman diagrams for the QED Compton process (QEDCS). $X \equiv p$
(and $P_X \equiv P'$) 
corresponds to elastic scattering.}}
\end{center}
\vspace{1cm}
\begin{center}
\epsfig{figure= 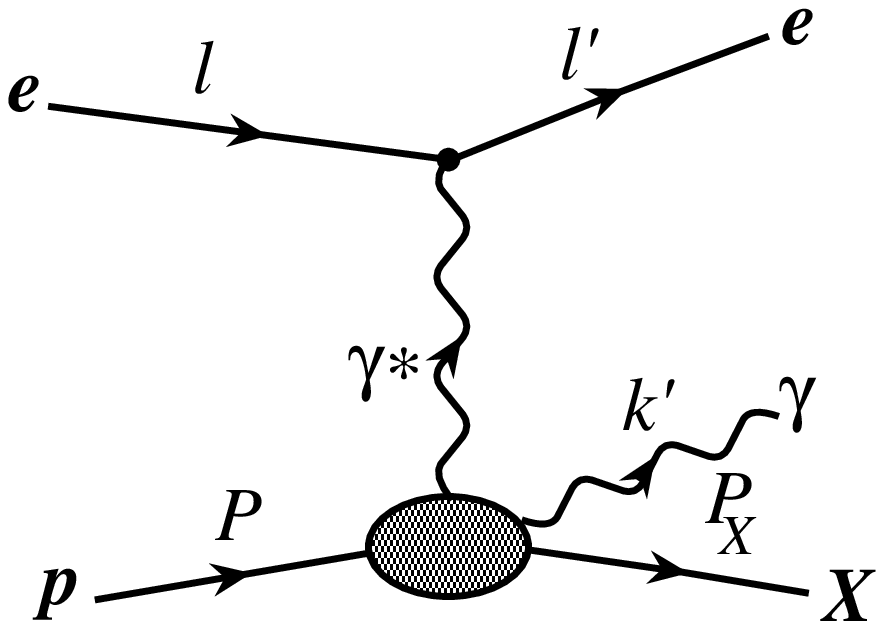, width=16cm, height= 6cm}\\
\end{center}
\begin{center}
\parbox{10.0cm}
{{\footnotesize
 Fig. 2:  As in Fig. 1 but for the virtual Compton scattering (VCS)
background process. }}
\end{center}

\newpage  
\begin{center}
\parbox{8cm}{\epsfig{figure=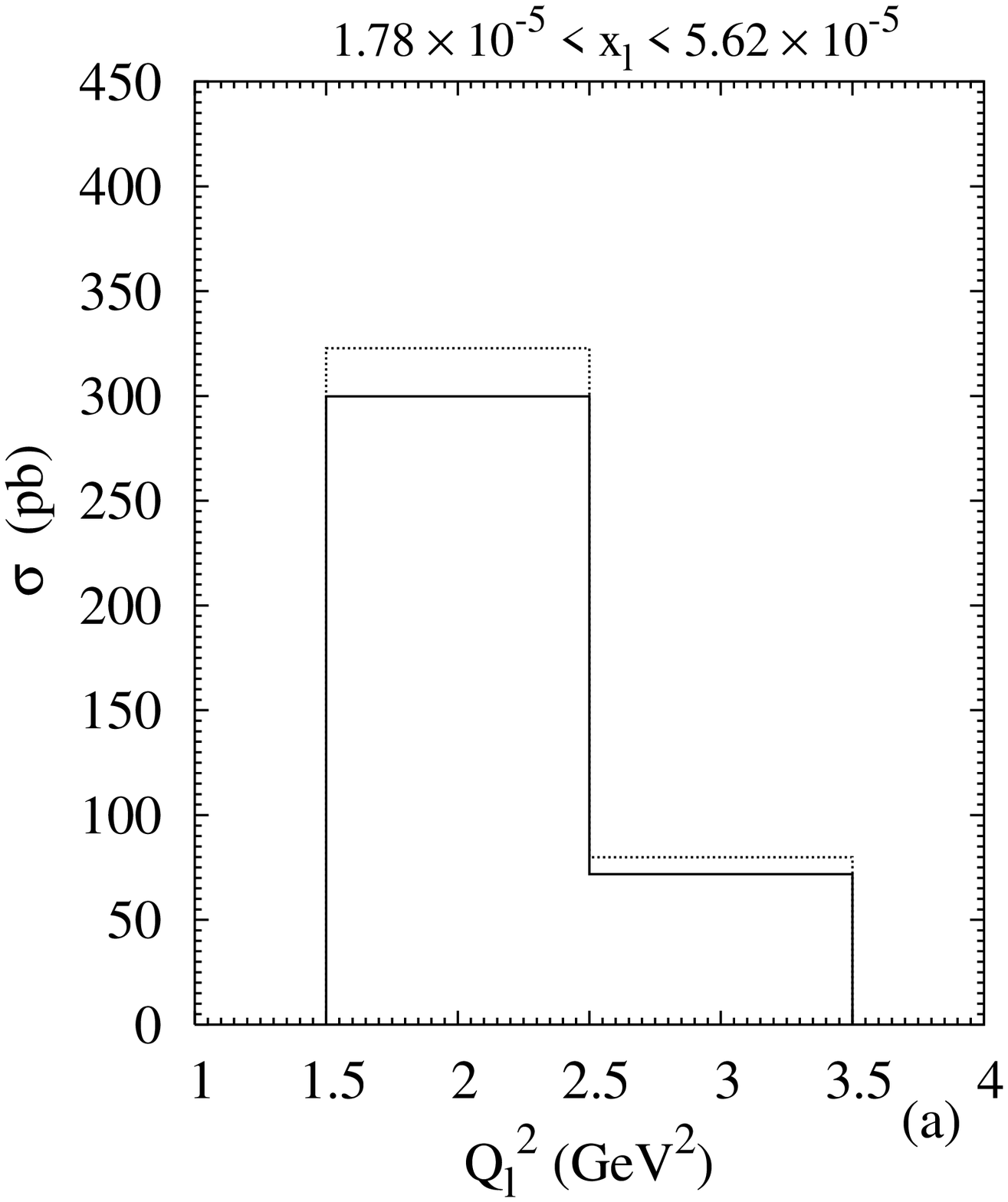,width=8.5 cm,height=7.5 cm}}\
\
\parbox{8cm}{\epsfig{figure=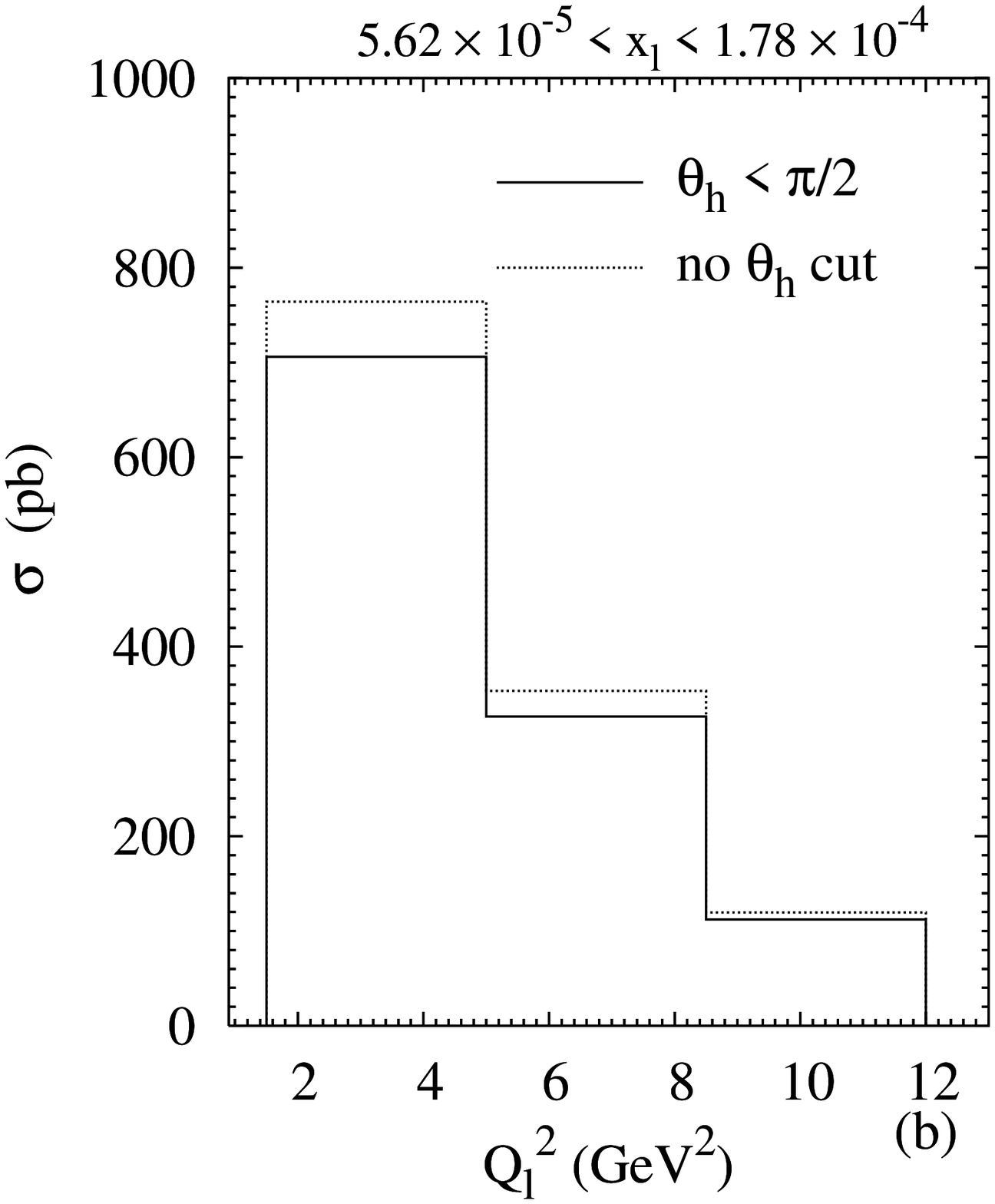,width=8.5 cm,height=7.5 cm}}\
\
\end{center}
\vspace{0.3cm}
\begin{center}
\hspace{-0.3cm}
\parbox{8cm}{\epsfig{figure=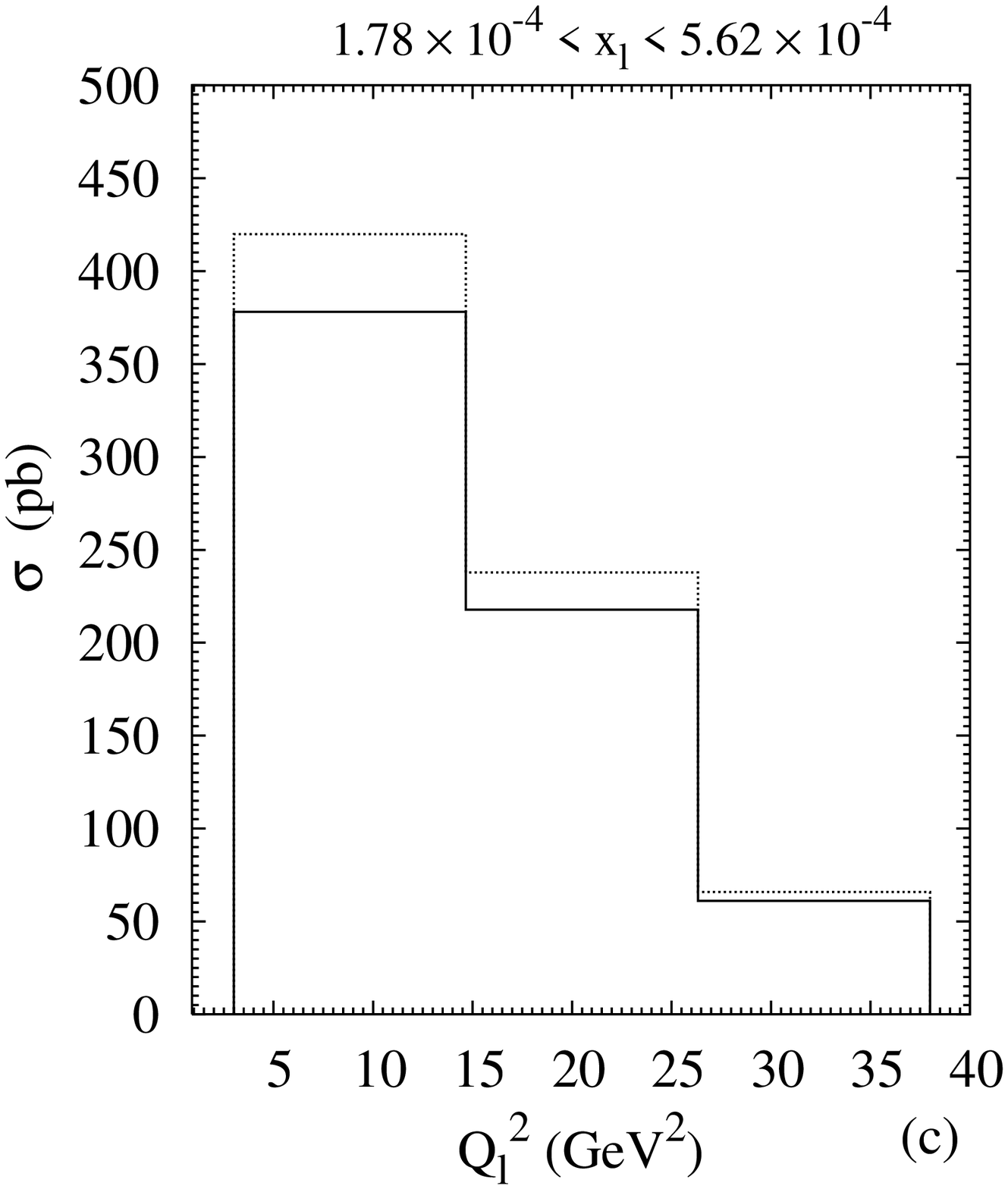,width=8.5 cm,height=7.5 cm}}\
\
\parbox{8cm}{\epsfig{figure=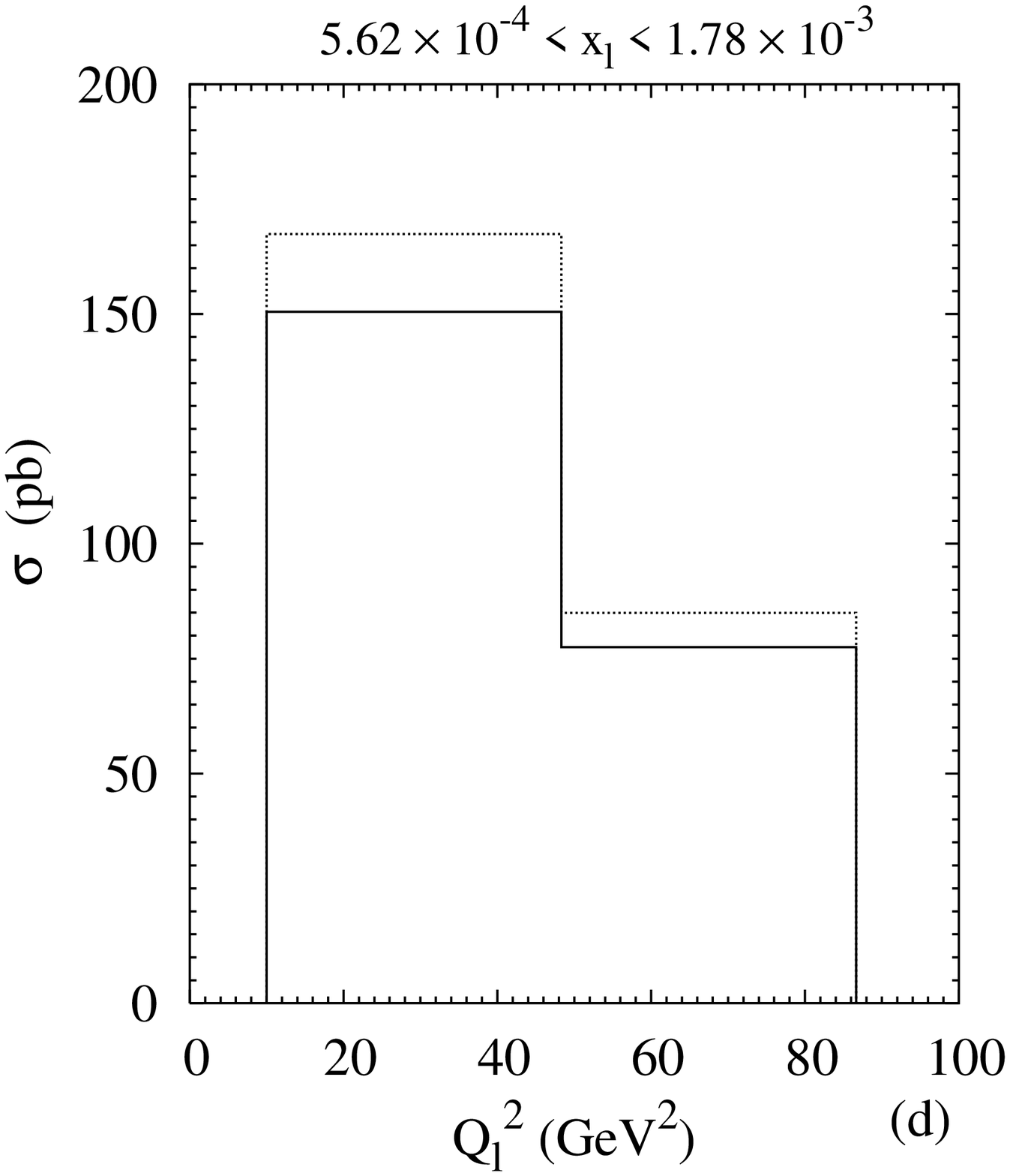,width=8.5 cm,height=7.5 cm}}\
\
\end{center}
\vspace{0.2cm}
\begin{center}
\parbox{14.0cm}
{{\footnotesize
 Fig. 3:  Double differential cross section for QED Compton scattering at
HERA-H1.  The kinematical bins correspond to Table 1 of \cite{pap1}. 
The continuous line describes the total (elastic + inelastic) cross section 
subject to the set of cuts A in table I. The dotted line shows the same results
when the constraint on $\theta_h$ is removed.}}
\end{center}
\newpage
\vspace*{4cm}
\begin{center}
\epsfig{figure= 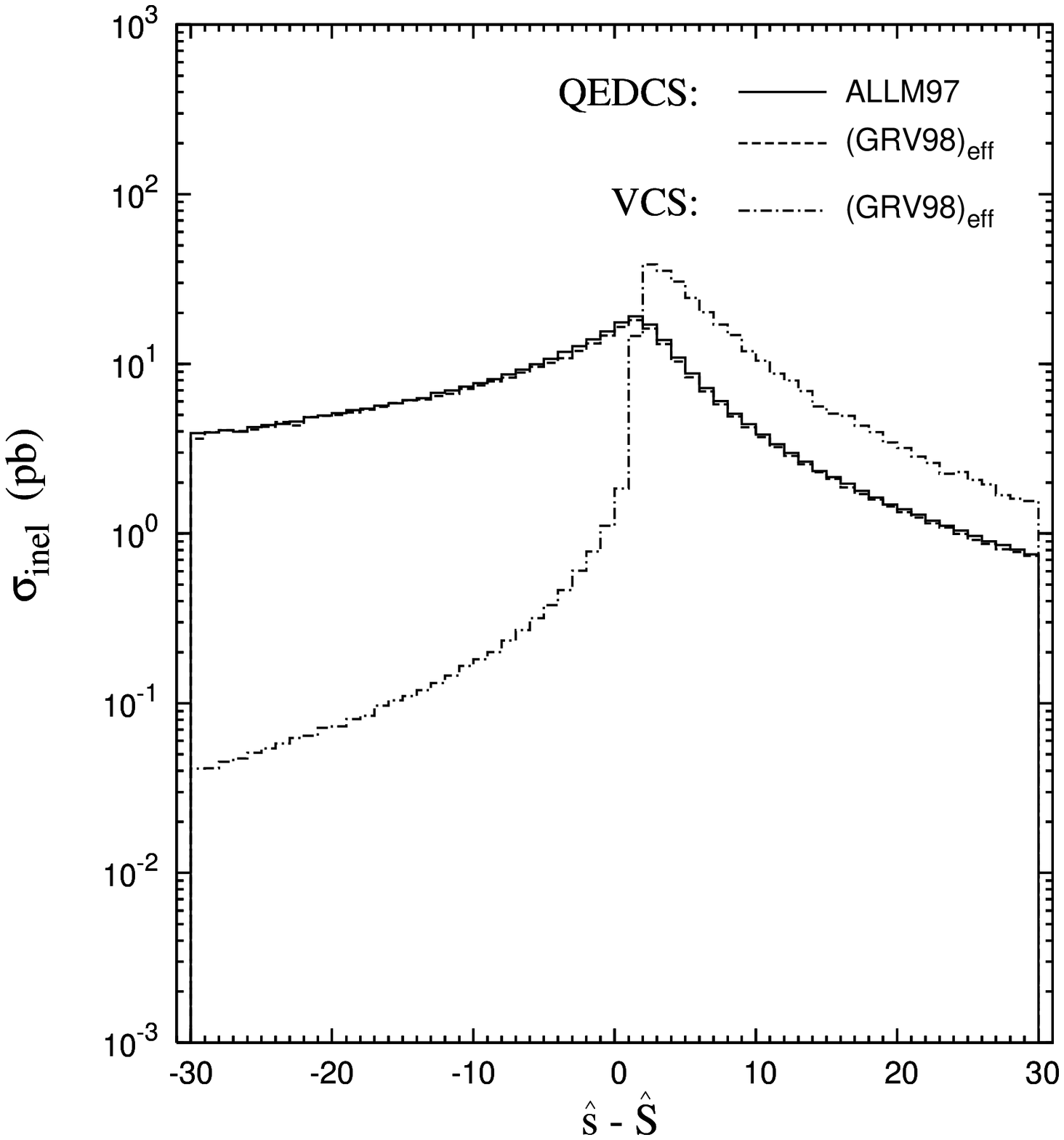,width=12.0cm,height=11.0cm}\\
\end{center}
\begin{center} 
\parbox{14.0cm}
{{\footnotesize
 Fig. 4:  Cross section for the QEDCS and VCS processes (inelastic) at HERA-H1. 
The bins are in $\hat s- \hat S$, expressed in $\mathrm{GeV^2}$. The cuts applied are listed in table I, set B (except $\hat S \gtrsim \hat s$). The continuous line corresponds to the QEDCS cross section with
ALLM97 parametrization of $F_2(x_B, Q^2)$, the dashed line corresponds to
the  
QEDCS cross section using the effective GRV98 parton distributions in 
Eq. (\ref{fq}) and  the dashed dotted line corresponds to the
VCS cross section using the same effective distributions.}}
\end{center}
\newpage
\vspace*{4cm}
\begin{center}
\parbox{8cm}{\epsfig{figure=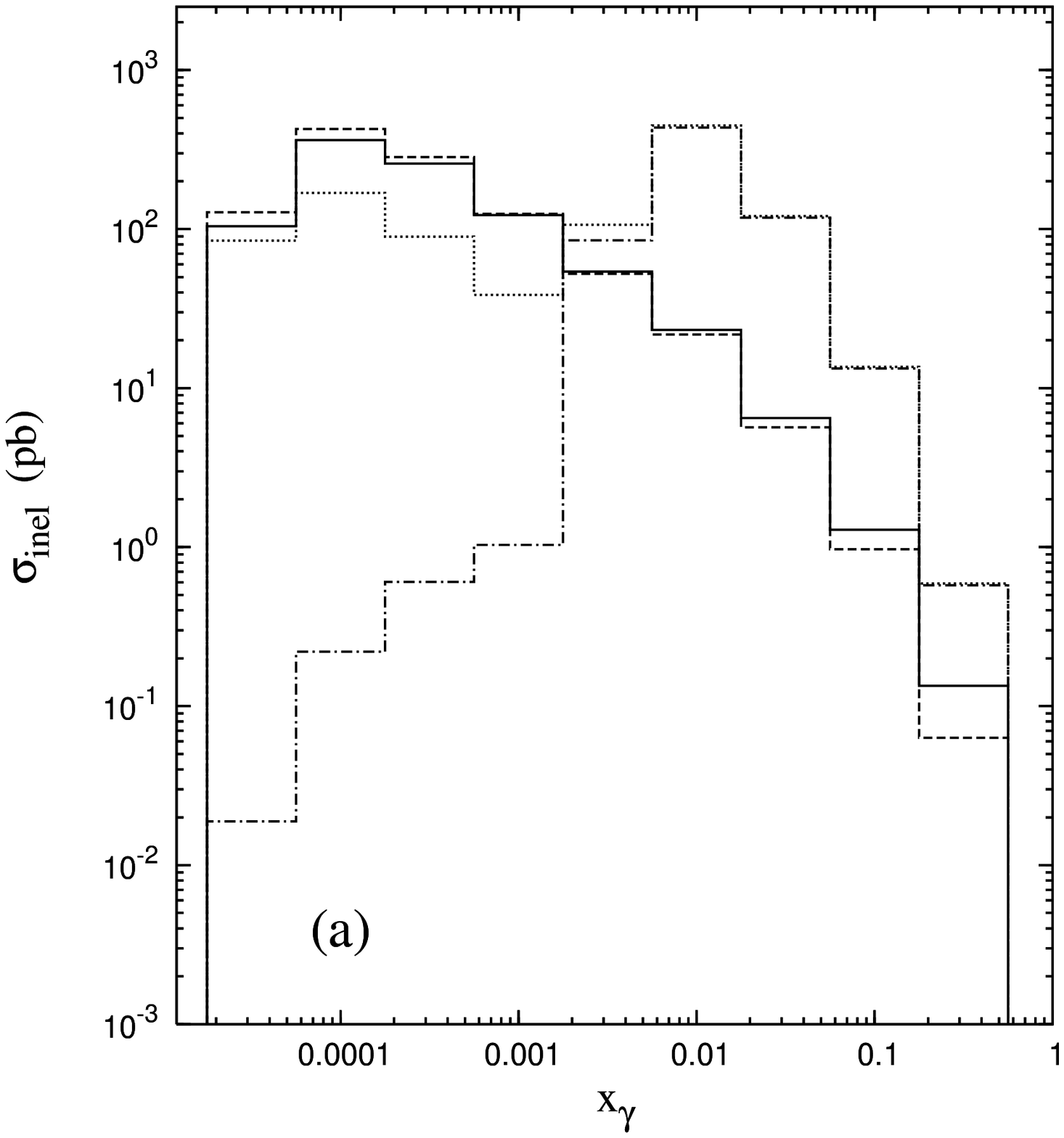,width=8.5 cm,height=7.5 cm}}\
\
\parbox{8cm}{\epsfig{figure=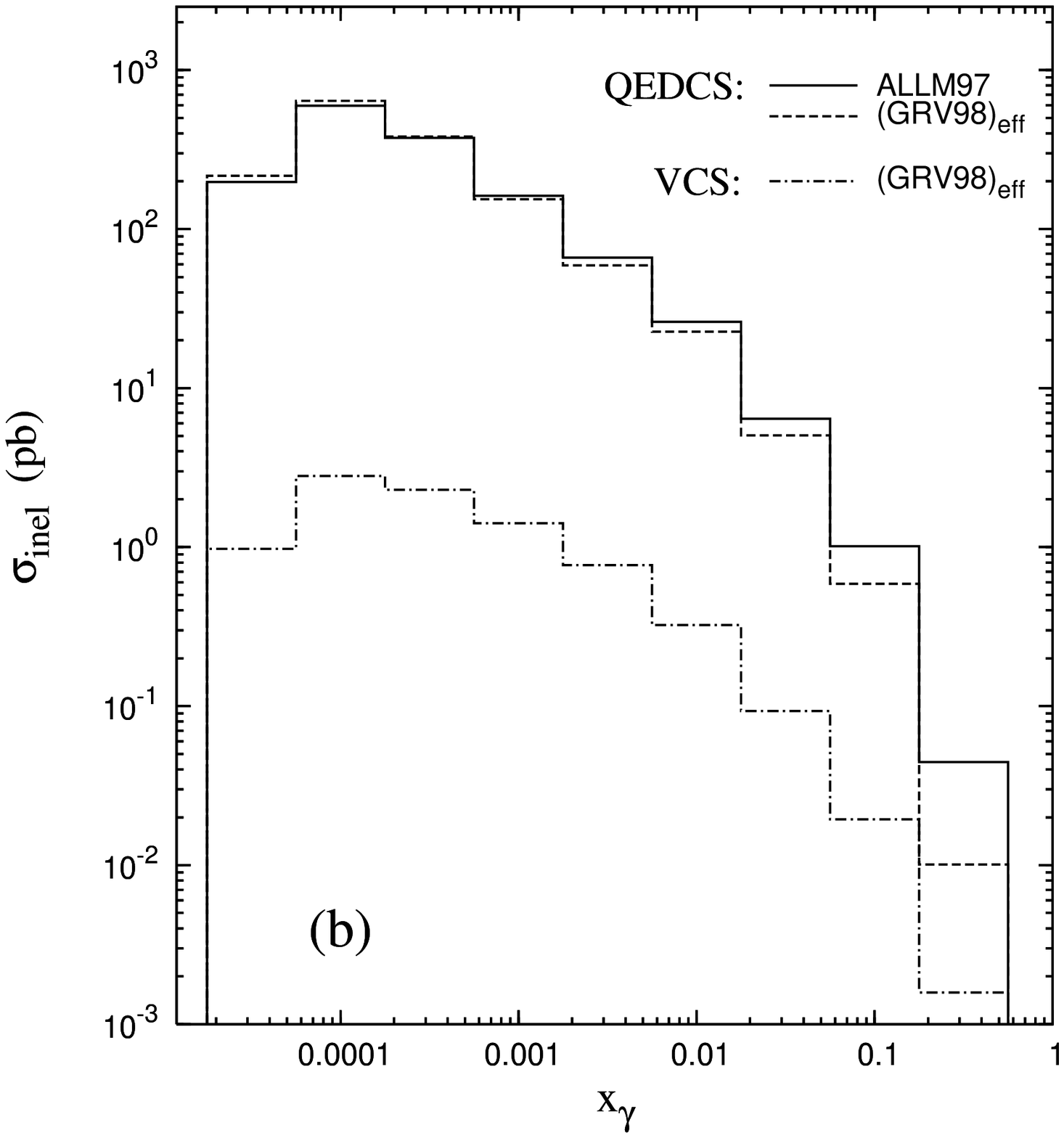,width=8.5 cm,height=7.5 cm}}\
\
\end{center}
\vspace{0.2cm}
\begin{center}
\parbox{14.0cm}
{{\footnotesize
Fig. 5: Cross section for QED Compton scattering in bins of $x_{\gamma}$
as calculated with the ALLM97 (full line) and the 
$\mathrm{(GRV98)}_{\mathrm{eff}}$ (dashed line) parametrization of 
$F_2(x_B, Q^2)$, respectively,  as compared to the VCS background cross section (dot-dashed line). The cuts employed are: a) as in set A, b) as in set B of 
table I. The dotted line in Fig. 5 a) shows the VCS cross section subject to
the set of cuts A without the constraint on $\theta_h$.}}
\end{center}
\newpage
\vspace*{4cm}
\begin{center}
\parbox{8cm}{\epsfig{figure=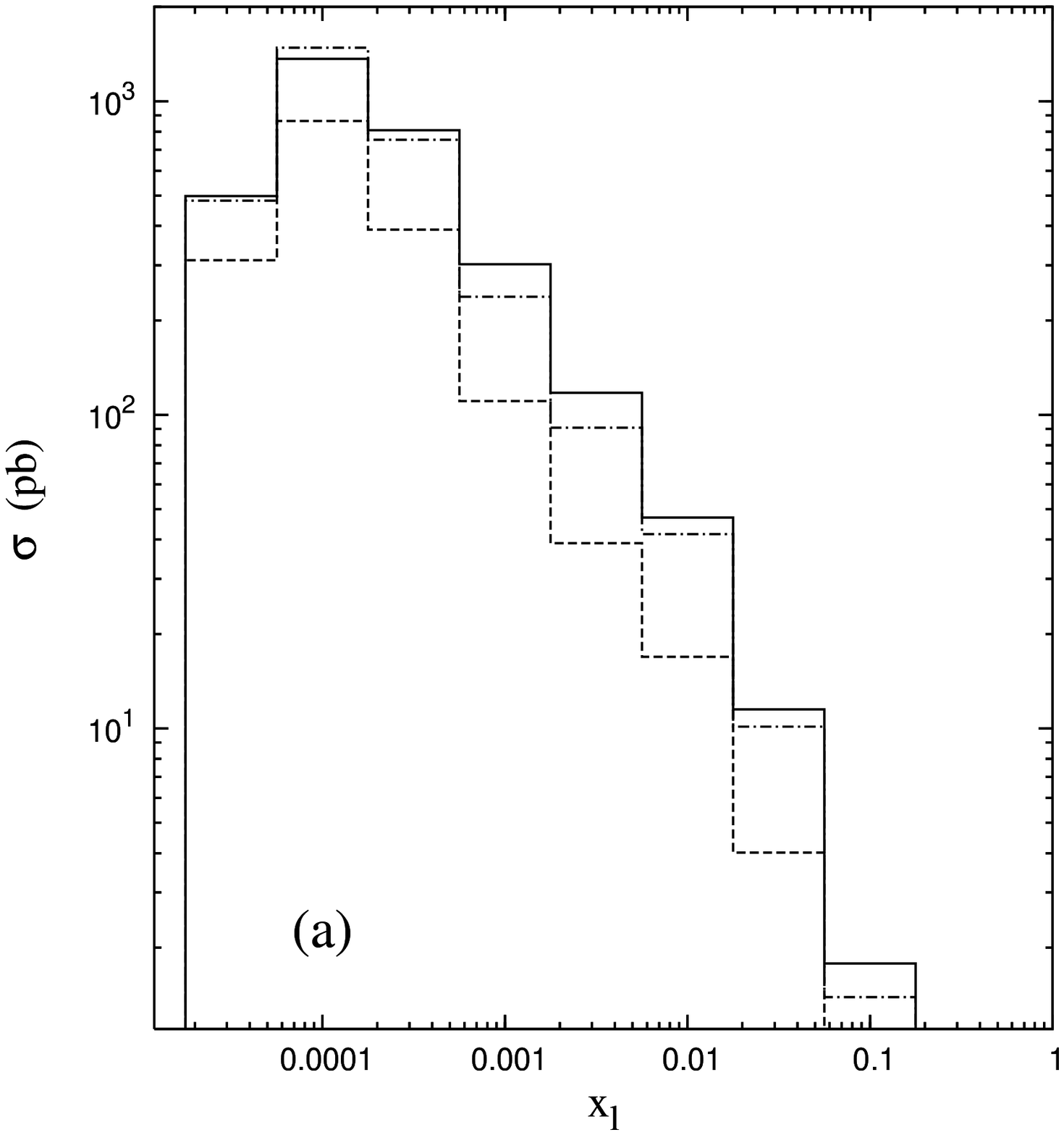,width=8.5 cm,height=7.5 cm}}\
\
\parbox{8cm}{\epsfig{figure=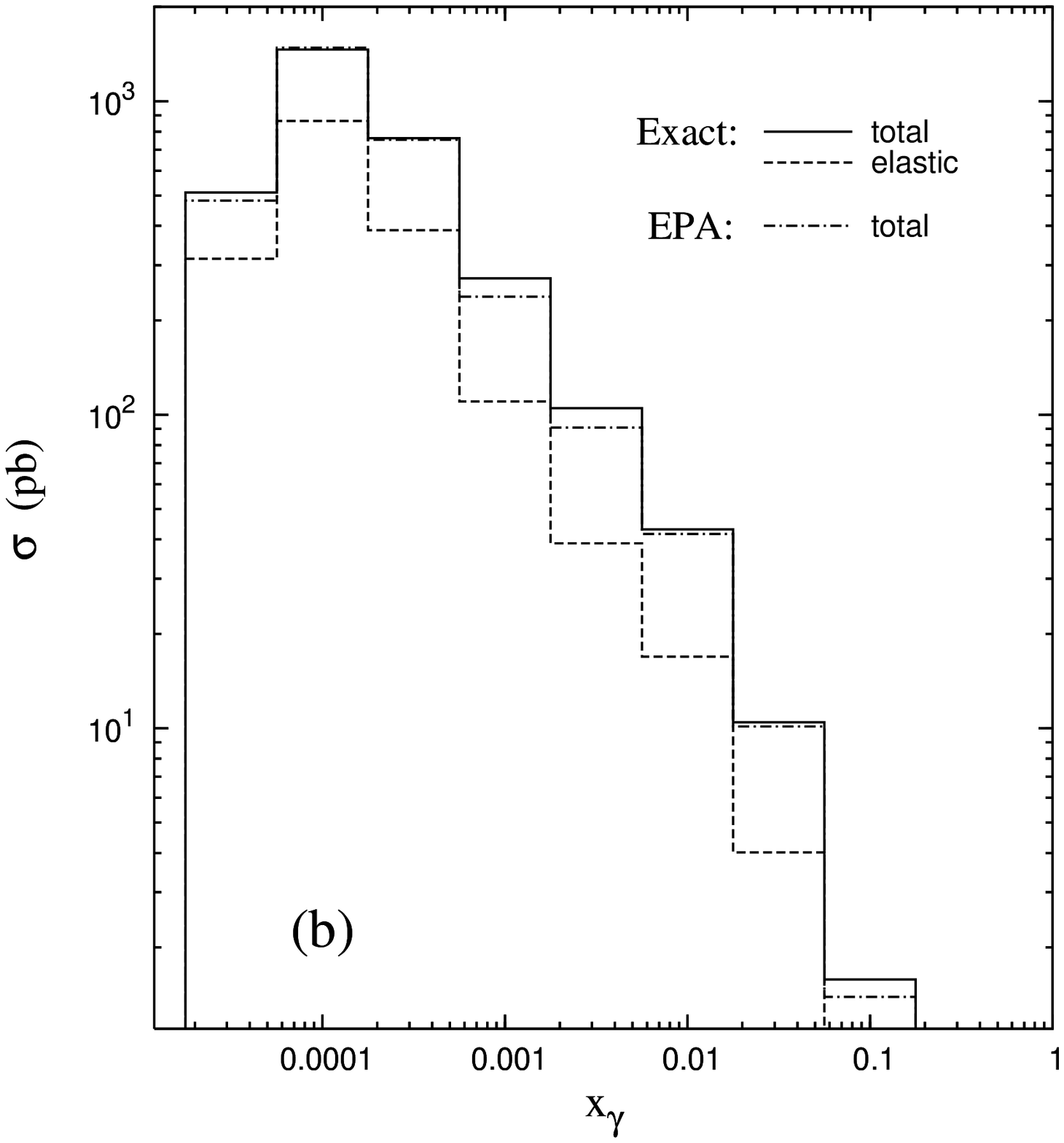,width=8.5 cm,height=7.5 cm}}\
\
\end{center}
\vspace{0.2cm}
\begin{center}
\parbox{14.0cm}
{{\footnotesize
Fig. 6:  Cross section for QED Compton scattering at HERA-H1 subject to the 
cuts of set B in table I, in (a) $x_l$ bins, (b) $x_{\gamma}$ bins. 
The continuous line corresponds to our exact calculation using ALLM97 
parametrization of $F_2(x_B, Q^2)$, the dot-dashed line corresponds to the 
same in the EPA, the dashed line shows the elastic contribution. }}
\end{center}

\begin{thebibliography}{99} 
\bibitem{kessler} A. Courau and P. Kessler, Phys. Rev. {\bf D 46}, 117,
(1992).   
\bibitem{blu} J. Bl\"umlein, G. Levman, H. Spiesberger, J. Phys. {\bf G 19},
1695 (1993). 
\bibitem{rujula} A. De Rujula, W. Vogelsang, Phys. Lett. {\bf B 451} 437
(1999). 
\bibitem{kniehl} B. Kniehl, Phys. Lett. {\bf B 254}, 267, (1991).
\bibitem{drees} M. Drees, R. M. Godbole, M. Nowakowski, S. Rindani, Phys.
Rev. {\bf D 50}, 2335 (1994).
\bibitem{ohn} J. Ohnemus, T. F. Walsh, P. M. Zerwas, Phys. Lett. {\bf B
328}, 369 (1994).  
\bibitem{gsv} M. Gl\"uck, M. Stratmann, W. Vogelsang, Phys. Lett. {\bf B
343}, 399 (1995).
\bibitem{gpr1} M. Gl\"uck, C. Pisano, E. Reya, Phys. Lett. {\bf B 540}, 75,
(2002).  
\bibitem{gpr2} M. Gl\"uck, C. Pisano, E. Reya, I. Schienbein,
Eur. Phys. J. {\bf C 27}, 427 (2003).
\bibitem{lend} V. Lendermann, H. C. Schultz-Coulon, D. Wegener, Eur. Phys.
Jour. {\bf C 31} 343 (2003).
\bibitem{thesis} V. Lenderman, Ph. D. thesis, Univ. Dortmund,
H1 collaboration, DESY-THESIS-2002-004, (2002).
\bibitem{pap1} A. Mukherjee and C. Pisano, Eur. Phys. J. {\bf C 30}, 477
(2003).
\bibitem{ji} X. Ji, Phys. Rev. {\bf D 55}, 7114 (1997).
\bibitem{allm} H. Abramowicz and A. Levy, hep-ph/9712415.
\bibitem{brodsky} S. J. Brodsky, J. F. Gunion, R. Jaffe, Phys. Rev. {\bf D
6}, 2487 (1972).
\bibitem{metz} P. Hoyer, M. Maul and A. Metz, Eur. Phys. J. {\bf C 17}, 113 
(2000).
\bibitem{grv} M. Gl\"uck, E. Reya, A. Vogt, Eur. Phys. J. {\bf
C 5}, 461 (1998). 
\bibitem{bad} B. Badelek and J. Kwiecinski, Phys. Lett. {\bf B 295}, 263 (1992).
\end{thebibliography}
\end{document}